\documentclass[preprint,prb,aps,a4paper,epsf,draft,showkeys,superscriptaddress] {revtex4} 
\usepackage[final]{graphicx}   
\usepackage{dcolumn}           
\usepackage{bm}                    
\usepackage{dcolumn}           
\usepackage{longtable} 
\usepackage{rotating}
\usepackage{color}
\def\abinitio{{\it ab initio}}

\def\cmm1{cm$^{-1}$}

\def\YAG{Y$_3$Al$_5$O$_{12}$}
\def\LuCaMgSiG{Lu$_2$CaMg$_2$Si$_3$O$_{12}$}
\def\CaScSiG{Ca$_3$Sc$_2$Si$_3$O$_{12}$}

\def\CeIII{Ce$^{3+}$}

\def\CeOviiimxiii{(CeO$_8$)$^{13-}$}

\def\OCeiv+xiv{(OCe$_4$)$^{14+}$}
\def\OUiv+xiv{(OU$_4$)$^{14+}$}

\begin{document}
\title{ 
       $4f$ and $5d$ levels of  Ce$^{3+}$ in $D_2$ eightfold oxygen coordination
      }
\date{\today}
\author{Luis Seijo}
\thanks{Corresponding author}
\email{luis.seijo@uam.es}
\affiliation{Departamento de Qu\'{\i}mica, 
             Universidad Aut\'onoma de Madrid, 28049 Madrid, Spain}
\affiliation{Instituto Universitario de Ciencia de Materiales Nicol\'as Cabrera,
             Universidad Aut\'onoma de Madrid, 28049 Madrid, Spain}
\author{Zoila Barandiar\'an}
\affiliation{Departamento de Qu\'{\i}mica, 
             Universidad Aut\'onoma de Madrid, 28049 Madrid, Spain}
\affiliation{Instituto Universitario de Ciencia de Materiales Nicol\'as Cabrera,
             Universidad Aut\'onoma de Madrid, 28049 Madrid, Spain}
\begin{abstract}
The effects 
on the $4f$ and $5d$ levels of \CeIII\ 
of its first coordination shell geometry in Ce-doped oxides with a $D_2$  8-fold  site,
like  garnets, 
are studied with embedded cluster, wave function based \abinitio\ methods.
The only deformations of a CeO$_8$ cube 
that are found to shift the lowest $4f \rightarrow 5d$ transition to the red
are the symmetric Ce-O bond compression
and the tetragonal symmetric bond bending.
These results are analyzed in terms of centroid and ligand field stabilization energy differences.
The splittings of the upper $5d$ levels and of the $4f$ levels are also  discussed.
\end{abstract}
\maketitle
\section{\label{SEC:intro}Introduction}

Yttrium aluminum garnet \YAG\ (YAG) doped with \CeIII\ is a well known phosphor
with a blue  \CeIII\ $4f \rightarrow 5d$ absorption and a corresponding yellow $5d \rightarrow 4f$ emisssion,~\cite{BLASSE:67}
which is used in InGaN blue LED based, white light solid-state lighting devices.~\cite{NAKAMURA:97,JUSTEL:98}
In the search for alternative phosphors with the best efficiencies and tailored color-rendering indexes, 
doping \CeIII\ in other artificial garnets has been a  line of work.
It led, for instance, 
to the discoveries of 
the \LuCaMgSiG:\CeIII\ orange phosphor~\cite{SETLUR:06} and 
the \CaScSiG:\CeIII\ green phosphor.~\cite{SHIMOMURA:07}

Most garnets can be described in terms of
a 160 atom body-centered cubic unit cell (80 atom primitive cell) of the $Ia\bar{3}d$ (230) space group,
which contains eight formula units of A$_{3}$B$_{2}'$B$_{3}''$O$_{12}$.
A, B$'$ and B$''$ are cations of different nominal charges  in different symmetry sites.
In \YAG, for instance,
A$\equiv$Y, B$'$$\equiv$Al, and B$''$$\equiv$Al,~\cite{EULER:65} 
all with oxidation states +3.
In \CaScSiG,
A$\equiv$Ca, B$'$$\equiv$Sc, and B$''$$\equiv$Si,~\cite{QUARTIERI:06} 
with respective oxidation sates +2, +3, and +4.
In \LuCaMgSiG,
2/3 of the A sites are occupied by Lu and 1/3 by Ca atoms,~\cite{SETLUR:06} 
and B$'$$\equiv$Mg, and B$''$$\equiv$Si, 
with respective oxidation states +3, +2, +2 and +4.
In all cases, 
A occupy 24(c) sites of 8-fold coordination, 
B$'$ 16(a) sites of 6-fold coordination, 
and B$''$ 24(d) sites of 4-fold coordination, 
all of them at fixed positions,
with the remaining 96 O atoms in (h) sites, which depend on three $x$, $y$ and $z$ internal parameters.~\cite{EULER:65} 
Optically active \CeIII\ impurities substitute for A atoms at (c) sites, 
which have the local symmetry of the $D_2$ point  group
and a first coordination shell made of 8 oxygen atoms.


The energies of the $4f$ and $5d$ levels of \CeIII\ in particular garnets and in other oxides depend on 
the bonding and electrostatic interactions between Ce and the hosts,  and 
they can be calculated in a one-by-one basis, with reasonable accuracy, by means of \abinitio\ methods,
both in $D_2$ local symmetry like in Ce-doped YAG~\cite{GRACIA:08}
and in lower local symmetries like in Ce,La-doped YAG~\cite{MUNOZ-GARCIA:10:b} and Ce,Ga-doped YAG.~\cite{MUNOZ-GARCIA:10:a}
These energies are dominated, up to first-order, by the bonding interactions between Ce and its first oxygen coordination shell, subject to the basic confinement embedding effects of the host.
Other specific host embedding effects are important at higher orders of approximation and they refine the results.

This paper is aimed at providing a basic study of the energies of the levels of the $4f^1$ and $5d^1$ manifolds of \CeIII\ in $D_2$  8-fold oxygen coordination.
We report the results of an \abinitio\ calculation of them after consideration of the bonding interactions with the first oxygen coordination shell only, under the effects of a cubic, non host specific, confinement embedding potential.
The present results are a common reference for these levels of \CeIII\ in specific garnets 
(and other oxides with 8-fold coordination),
since the latter can be considered as resulting from the former after including the host specific embedding effects.

\section{\label{SEC:method}Method}

In order to know the $4f^1$ and $5d^1$ energy levels of  \CeIII\
under the effects of its interactions with a coordination shell of eight oxygens
in a $D_2$ symmetry site of a solid oxide,
we calculated the corresponding energy levels of the \CeOviiimxiii\ cluster
embedded in a $O_h$ host potential. 
We will describe later the details of the \abinitio\ wave function based quantum mechanical calculation.
With he choice of a cubic host embedding potential, 
all the energy splittings and changes due to  $D_2$ distortions are ascribable to the interactions
between Ce and the first-shell O atoms. 

It is convenient to look at the $D_2$ energy levels as derived from the more familiar $O_h$ levels
after geometry distortions and this is what we did in this work. 
In consequence with this, we decided to take a cubic  CeO$_8$ $O_h$ moiety as a reference
and to chose six atomic displacement coordinates $S_1 - S_6$
that transform according to irreducible representations
of the $O_h$ point symmetry group
for the six degrees of freedom of the CeO$_8$ $D_2$ moiety.
They are defined in Table~\ref{TAB:S1-S6} and represented in Fig.~\ref{FIG:S1-S6}.
$S_1$ represents the breathing (or symmetric bond stretching) of the reference cube and 
it transforms according to the totally symmetric irreducible representation $a_{1g}$;
$S_2$ represents an asymmetric bond stretching of
the crosses $a_1-a_2-a_3-a_4$ and $e_1-e_3-e_2-e_4$
and it transforms according to the $e_{g}\epsilon$ sub-especies of the doubly degenerate $e_{g}$ irreducible representation;~\cite{SUGANO:70}
$S_3$ and $S_4$ are the $e_{g}\theta$ and $e_{g}\epsilon$ symmetric and asymmetric bond bendings of the $a$ and $e$ crosses; 
and $S_5$ and $S_6$ are the $e_{u}\theta$ and $e_{u}\epsilon$ symmetric and asymmetric  twistings of the crosses.

The details of the quantum mechanical calculation are the following.
We performed \abinitio\ calculations on a \CeOviiimxiii\  cluster
that include Ce-O bonding effects, static and dynamic electron correlation effects, 
scalar and spin-orbit coupling relativistic effects, and cubic host embedding effects.
For each $D_2$ nuclear configuration of the  \CeOviiimxiii\ embedded cluster,
we performed, in a first step,
scalar relativistic calculations with the many-electron second-order Douglas-Kroll-Hess (DKH) Hamiltonian.~\cite{DOUGLAS:74,HESS:86}
These were
state-average complete-active-space self-consistent-field~\cite{ROOS:80,SIEGBAHN:80,SIEGBAHN:81} (SA-CASSCF) calculations 
with the active space that results from distributing the open-shell electron in 13 active molecular orbitals
with main character Ce $4f, 5d, 6s$, 
which provided occupied and empty molecular orbitals to feed subsequent 
multi-state second-order perturbation theory calculations (MS-CASPT2),~\cite{ANDERSSON:90,ANDERSSON:92,ZAITSEVSKII:95,FINLEY:98} 
where the dynamic correlation of 73 electrons 
(the $5s, 5p, 4f, 5d, 6s$ electrons of Ce and $2s, 2p$ electrons of the O atoms)
 were taken into account.
 In a second step, 
 spin-orbit coupling effects were included by adding the 
 AMFI approximation of the DKH spin-orbit coupling operator~\cite{HESS:96} to the Hamiltonian
 and performing restricted-active-space state-interaction spin-orbit calculations (RASSI-SO)~\cite{MALMQVIST:02}
 with the previously computed SA-CASSCF wave functions and MS-CASPT2 energies.
 
 In all these calculations, the Hamiltonian of the \CeOviiimxiii\  cluster is supplemented with
 a cubic embedding Hamiltonian. 
 For this we chose the \abinitio\ model potential embedding Hamiltonian (AIMP)~\cite{BARANDIARAN:88}
 of a SrO host with CsCl lattice structure ($Pm\overline{3}m$, no. 221),
 which provides a perfect 8-fold cubic oxide coordination of the cationic site.
 We used a lattice constant $a=$~2.87~\AA\ (the one that gives a Ce-O equilibrium distance in 
 the cubic \CeOviiimxiii\ embedded cluster of 2.34~\AA\ at the MS-CASPT2 level of calculation,
 which is the Ce-O distance of the reference CeO$_8$ cube in an hypothetical undistorted Ce-doped YAG~\cite{EULER:65}).
 The embedding potential of this  8-fold coordinated cubic oxide was computed in Hartree-Fock
 self-consistent embedded-ions calculations (SCEI),~\cite{SEIJO:91}
 in which the input embedding AIMPs used for the Sr$^{2+}$ and O$^{2-}$ ions of SrO are consistent
 with the output AIMPs obtained out of the HF orbitals of the embedded Sr$^{2+}$ and O$^{2-}$ ions.

\section{\label{SEC:results}Results}
The dependence of the $4f \rightarrow 4f$ and $4f \rightarrow 5d$ transitions 
on the $D_2$ distortions of the first coordination shell are shown 
in Fig.~\ref{FIG:S135S246-spinfree} without spin-orbit coupling and
in Fig.~\ref{FIG:S135S246-spinorbit} with spin-orbit coupling interactions included.
The chosen ranges of the $S_1 - S_6$ coordinates span their usual values in natural and artificial garnets,
which are shown in Table~\ref{TAB:garnets} and in the figures.
Besides the transitions, the figures include two additional lines: 
One is the difference between the $5d$ and $4f$ energy centroids (or average energies of the five $5d^1$ levels and the seven $4f^1$ levels), 
$\Delta E_{centroid,fd} = E_{centroid}(5d^1) - E_{centroid}(4f^1)$;
the other is the difference between the ligand field stabilization energies of the  $5d$ and $4f$  lowest levels
$\Delta E_{ligand\,field,df} = [ E_{centroid}(5d^1) - E_1(5d^1) ] - [ E_{centroid}(4f^1) - E_1(4f^1) ] $.
The fact that the lowest $4f \rightarrow 5d$ transition equals
their subtraction, $E_1(5d^1) - E_1(4f^1) = \Delta E_{centroid,fd}  - \Delta E_{ligand\,field,df} $,
has been used to analyze the reasons behind red and blue shifts of this transition.~\cite{MUNOZ-GARCIA:10:a,MUNOZ-GARCIA:10:b}

Let us first focus on the lowest $4f \rightarrow 5d$ transition,
whose reverse is the luminescence of \CeIII\ in garnets and other oxides.
It is clear that only two coordinates have an important impact on it:
the breathing mode $S_1$, 
which does not distort the cubic symmetry,
and the symmetric bond bending mode $S_3$, 
which gives a $e_g\theta$ tetragonal  distortion of the CeO$_8$ cube.

Symmetric compression of the Ce-O bonds lowers the  $4f \rightarrow 5d$ transition.
This is almost entirely due to the increase of the $5d$ ligand field it produces, 
which lowers the first $5d^1$ level with respect to its centroid.
This effect is slightly compensated with a small increase due to the centroids:
bond compression slightly shifts  the $5d$ centroid upwards with respect to the $4f$ centroid.
The latter observation has been made before~\cite{BARANDIARAN:05:a,MUNOZ-GARCIA:10:a}
and it contradicts the predictions of the usual model for the centroid energy difference,~\cite{JUDD:77,MORRISON:80:a,BETTINELLI:01}
according to which it should be smaller for smaller bond lengths;
nevertheless, the model has been useful to rationalize $5d \rightarrow 4f$ luminescences.~\cite{DORENBOS:00}
The contribution of the breathing mode $S_1$ to the lowest $4f \rightarrow 5d$ transition 
represented in Fig.~\ref{FIG:S135S246-spinorbit}
fits $8255\,S_1 - 4803\,S_1^2$;
for cubic CeO$_8$ moieties $S_1 = 2 \sqrt 2 \Delta d$, 
$\Delta d$ being the Ce-O bond length change with respect to the reference cube;
in our case the reference cube has Ce-O bond length of 2.34~\AA\ and a lowest $4f \rightarrow 5d$ transition of 26000~\cmm1.

Symmetric bond bending along the $S_3$ coordinate lowers de  $4f \rightarrow 5d$ transition.
Among the $D_2$ distortions of the cube, it is the only tetragonal one.  
Also, it is the only one  with a significant impact on the transition
(for the ranges of $S_2 - S_6$ shown by garnets).
The enhancement of tetragonal distortions of the cube as responsible for the red-shifts of the \CeIII\
luminescence has been suggested by Cheetham and coworkers~\cite{WU:07:b} 
and it is supported by the present investigation.
The image emerging from an analysis in terms of configurational energy centroids and ligand field stabilization
is one in which the lowering of the first $5d^1$ level with respect to the ground state is due
in almost equal amounts to an increase of the ligand field (of its tetragonal component) 
and to a reduction of the centroid energy difference.
Again, the latter effect contradicts the prediction of the Judd-Morrison model:
a very small increase of the centroid energy difference out of a very small increase of the Ce-O distances.
The contribution of the tetragonal mode $S_3$ to the lowest $4f \rightarrow 5d$ transition 
represented in Fig.~\ref{FIG:S135S246-spinorbit}
fits $1844\,S_3 - 2885\,S_3^2$ for $S_3 < 0$ 
and $-1876\,S_3 - 141\,S_3^2$ for $S_3 > 0$.
This contribution is due in almost equal amounts to the lowering of the centroid energy difference
and the increasing of the ligand field.
The centroid effect fits $10\,S_3 - 2695\,S_3^2$ for $S_3 < 0$
and $12\,S_3 - 2581\,S_3^2$ for $S_3 > 0$;
the ligand field effect fits $1834\,S_3 - 190\,S_3^2$ for $S_3 < 0$
and $-1887\,S_3 + 2440\,S_3^2$ for $S_3 > 0$.

Apart from $S_1$ and $S_3$, only the asymmetric stretching $S_2$ has an effect, although very small,  
on the first  $4f \rightarrow 5d$ transition (it fits $-2033\,S_2^2$ in the range of the figures).

Regarding higher $4f \rightarrow 5d$ transitions,
several remarks can be made from 
Figs.~\ref{FIG:S135S246-spinfree} and \ref{FIG:S135S246-spinorbit}.
The gap between the first and the second $5d$ levels related with the cubic $^2E_g$
is due to the symmetric bending $S_3$ and to the symmetric twisting $S_5$,
with a significantly higher contribution from the latter in garnets; however,
 different splittings in different garnets would be due to $S_3$.
The splitting of the three upper levels, mostly related with the cubic $^2T_{2g}$,
comes basically from the symmetric bending $S_3$ and the asymmetric stretching $S_2$, 
with a small contribution from the symmetric twisting $S_5$ and 
negligible contributions from the asymmetric bending and twisting $S_4$ and $S_6$. 
The larger $S_1$ and the absolute values of $S_2$ and $S_3$, the higher probability for the third $5d$ level
to appear below the conduction band of the garnet.

Finally, 
the $4f$ levels are shown with more detail in Figs.~\ref{FIG:S135S246-spinfree-4f} and \ref{FIG:S135S246-spinorbit-4f}, without and with spin-orbit coupling respectively.
These may be interesting because the $5d \rightarrow 4f$ emission is made of the superposition of the 
seven individual emissions and the full with and shape of the emission band depends, although not only,
on the relative positions of the $4f$ levels. 
Since the emission to the highest $4f$ level has a minor contribution 
to the full emission band shape,~\cite{GRACIA:08} 
we can have a predictive hint by looking at the six lowest $4f$ levels.
According to Fig.~\ref{FIG:S135S246-spinorbit-4f},
the width seems to be controlled by $S_2$ and $S_3$, so that the garnets with
higher absolute values of these two coordinates would tend to have wider emission bands
because of their higher $4f$ level separation. Of course, the width of each individual $5d \rightarrow 4f$ emission
will depend on the $5d$ and $4f$ equilibrium offsets.

\section{\label{SEC:conclusions}Conclusions}
The levels of the $4f^1$ and $5d^1$ manifolds of \CeIII\ in $D_2$  8-fold oxygen coordination
have been calculated \abinitio\ as a function of  $D_2$ deformation coordinates 
of a reference cube, with the goal of pinpointing the effects of the geometry of the first coordination shell.
A \CeOviiimxiii\ cluster under the effects of a  confinement AIMP embedding potential of cubic symmetry
has been used
at a level of calculation  all-electron DKH for the Hamiltonian and
SA-CASSCF/MS-CASTP2/RASSI-SO for the wave functions.
The results include bonding, correlation, scalar relativistic, spin-orbit coupling, and embedding interactions.

It is found that the lowest $4f \rightarrow 5d$ transition shifts significantly to the red
only as a consequence of symmetric Ce-O bond compression and 
tetragonal symmetric bond bending.

The increase of the $5d$ cubic ligand field dominates the effects of the bond compression,
whereas the effect of the tetragonal bond bending distortion is divided in almost equal amounts
between an increase of the $5d$ tetragonal ligand field and 
a reduction of the energy difference between the $5d$ and $4f$ energy centroids.

The splittings of the upper $5d$ levels and of the $4f$ levels have also been studied.
\acknowledgments
This work was partly supported by
a grant from Ministerio de Econom\'{\i}a y Competitivad, Spain
(Direcci\'on General de Investigaci\'on y Gesti\'on del Plan Nacional de I+D+I,
MAT2011-24586).
%

\begin{thebibliography}{37}
\expandafter\ifx\csname natexlab\endcsname\relax\def\natexlab#1{#1}\fi
\expandafter\ifx\csname bibnamefont\endcsname\relax
  \def\bibnamefont#1{#1}\fi
\expandafter\ifx\csname bibfnamefont\endcsname\relax
  \def\bibfnamefont#1{#1}\fi
\expandafter\ifx\csname citenamefont\endcsname\relax
  \def\citenamefont#1{#1}\fi
\expandafter\ifx\csname url\endcsname\relax
  \def\url#1{\texttt{#1}}\fi
\expandafter\ifx\csname urlprefix\endcsname\relax\def\urlprefix{URL }\fi
\providecommand{\bibinfo}[2]{#2}
\providecommand{\eprint}[2][]{\url{#2}}

\bibitem[{\citenamefont{Blasse and Bril}(1967)}]{BLASSE:67}
\bibinfo{author}{\bibfnamefont{G.}~\bibnamefont{Blasse}} \bibnamefont{and}
  \bibinfo{author}{\bibfnamefont{A.}~\bibnamefont{Bril}}, \bibinfo{journal}{J.\
  Chem.\ Phys.} \textbf{\bibinfo{volume}{47}}, \bibinfo{pages}{5139}
  (\bibinfo{year}{1967}).

\bibitem[{\citenamefont{Nakamura and Fasol}(1997)}]{NAKAMURA:97}
\bibinfo{author}{\bibfnamefont{S.}~\bibnamefont{Nakamura}} \bibnamefont{and}
  \bibinfo{author}{\bibfnamefont{G.}~\bibnamefont{Fasol}},
  \emph{\bibinfo{title}{The blue laser diode: GaN based light emitters and
  lasers}} (\bibinfo{publisher}{Springer}, \bibinfo{address}{Berlin},
  \bibinfo{year}{1997}).

\bibitem[{\citenamefont{J\"ustel et~al.}(1998)\citenamefont{J\"ustel, Nikol,
  and Ronda}}]{JUSTEL:98}
\bibinfo{author}{\bibfnamefont{T.}~\bibnamefont{J\"ustel}},
  \bibinfo{author}{\bibfnamefont{H.}~\bibnamefont{Nikol}}, \bibnamefont{and}
  \bibinfo{author}{\bibfnamefont{C.}~\bibnamefont{Ronda}},
  \bibinfo{journal}{Angew.\ Chem.,\ Int.\ Ed.} \textbf{\bibinfo{volume}{37}},
  \bibinfo{pages}{3084} (\bibinfo{year}{1998}).

\bibitem[{\citenamefont{Setlur et~al.}(2006)\citenamefont{Setlur, Heward, Gao,
  Srivastava, Chandran, and Shankar}}]{SETLUR:06}
\bibinfo{author}{\bibfnamefont{A.~A.} \bibnamefont{Setlur}},
  \bibinfo{author}{\bibfnamefont{W.~J.} \bibnamefont{Heward}},
  \bibinfo{author}{\bibfnamefont{Y.}~\bibnamefont{Gao}},
  \bibinfo{author}{\bibfnamefont{A.~M.} \bibnamefont{Srivastava}},
  \bibinfo{author}{\bibfnamefont{R.~G.} \bibnamefont{Chandran}},
  \bibnamefont{and} \bibinfo{author}{\bibfnamefont{M.~V.}
  \bibnamefont{Shankar}}, \bibinfo{journal}{Chem.\ Mater.}
  \textbf{\bibinfo{volume}{18}}, \bibinfo{pages}{3314} (\bibinfo{year}{2006}).

\bibitem[{\citenamefont{Shimomura et~al.}(2007)\citenamefont{Shimomura, Honma,
  Shigeiwa, Akai, Okamoto, and Kijima}}]{SHIMOMURA:07}
\bibinfo{author}{\bibfnamefont{Y.}~\bibnamefont{Shimomura}},
  \bibinfo{author}{\bibfnamefont{T.}~\bibnamefont{Honma}},
  \bibinfo{author}{\bibfnamefont{M.}~\bibnamefont{Shigeiwa}},
  \bibinfo{author}{\bibfnamefont{T.}~\bibnamefont{Akai}},
  \bibinfo{author}{\bibfnamefont{K.}~\bibnamefont{Okamoto}}, \bibnamefont{and}
  \bibinfo{author}{\bibfnamefont{N.}~\bibnamefont{Kijima}},
  \bibinfo{journal}{J.\ Electrochem.\ Soc.} \textbf{\bibinfo{volume}{154}},
  \bibinfo{pages}{J35} (\bibinfo{year}{2007}).

\bibitem[{\citenamefont{Euler and Bruce}(1965)}]{EULER:65}
\bibinfo{author}{\bibfnamefont{F.}~\bibnamefont{Euler}} \bibnamefont{and}
  \bibinfo{author}{\bibfnamefont{J.~A.} \bibnamefont{Bruce}},
  \bibinfo{journal}{Acta\ Crystallogr.} \textbf{\bibinfo{volume}{19}},
  \bibinfo{pages}{971} (\bibinfo{year}{1965}).

\bibitem[{\citenamefont{Quartieri et~al.}(2006)\citenamefont{Quartieri, Oberti,
  Boiocchi, Dalconi, Boscherini, Safonova, and Woodland}}]{QUARTIERI:06}
\bibinfo{author}{\bibfnamefont{S.}~\bibnamefont{Quartieri}},
  \bibinfo{author}{\bibfnamefont{R.}~\bibnamefont{Oberti}},
  \bibinfo{author}{\bibfnamefont{M.}~\bibnamefont{Boiocchi}},
  \bibinfo{author}{\bibfnamefont{M.~C.} \bibnamefont{Dalconi}},
  \bibinfo{author}{\bibfnamefont{F.}~\bibnamefont{Boscherini}},
  \bibinfo{author}{\bibfnamefont{O.}~\bibnamefont{Safonova}}, \bibnamefont{and}
  \bibinfo{author}{\bibfnamefont{A.~B.} \bibnamefont{Woodland}},
  \bibinfo{journal}{Am.\ Mineral.} \textbf{\bibinfo{volume}{91}},
  \bibinfo{pages}{1240} (\bibinfo{year}{2006}).

\bibitem[{\citenamefont{Gracia et~al.}(2008)\citenamefont{Gracia, Seijo,
  Barandiar\'an, Curulla, Niemansverdriet, and van Gennip}}]{GRACIA:08}
\bibinfo{author}{\bibfnamefont{J.}~\bibnamefont{Gracia}},
  \bibinfo{author}{\bibfnamefont{L.}~\bibnamefont{Seijo}},
  \bibinfo{author}{\bibfnamefont{Z.}~\bibnamefont{Barandiar\'an}},
  \bibinfo{author}{\bibfnamefont{D.}~\bibnamefont{Curulla}},
  \bibinfo{author}{\bibfnamefont{H.}~\bibnamefont{Niemansverdriet}},
  \bibnamefont{and} \bibinfo{author}{\bibfnamefont{W.}~\bibnamefont{van
  Gennip}}, \bibinfo{journal}{J.\ Lumin.} \textbf{\bibinfo{volume}{128}},
  \bibinfo{pages}{1248} (\bibinfo{year}{2008}).

\bibitem[{\citenamefont{{Mu\~noz-Garc\'{\i}a}
  et~al.}(2010)\citenamefont{{Mu\~noz-Garc\'{\i}a}, Pascual, Barandiar\'an, and
  Seijo}}]{MUNOZ-GARCIA:10:b}
\bibinfo{author}{\bibfnamefont{A.~B.} \bibnamefont{{Mu\~noz-Garc\'{\i}a}}},
  \bibinfo{author}{\bibfnamefont{J.~L.} \bibnamefont{Pascual}},
  \bibinfo{author}{\bibfnamefont{Z.}~\bibnamefont{Barandiar\'an}},
  \bibnamefont{and} \bibinfo{author}{\bibfnamefont{L.}~\bibnamefont{Seijo}},
  \bibinfo{journal}{Phys.\ Rev.\ B} \textbf{\bibinfo{volume}{82}},
  \bibinfo{pages}{064114} (\bibinfo{year}{2010}).

\bibitem[{\citenamefont{{Mu\~noz-Garc\'{\i}a} and
  Seijo}(2010)}]{MUNOZ-GARCIA:10:a}
\bibinfo{author}{\bibfnamefont{A.~B.} \bibnamefont{{Mu\~noz-Garc\'{\i}a}}}
  \bibnamefont{and} \bibinfo{author}{\bibfnamefont{L.}~\bibnamefont{Seijo}},
  \bibinfo{journal}{Phys.\ Rev.\ B} \textbf{\bibinfo{volume}{82}},
  \bibinfo{pages}{184118} (\bibinfo{year}{2010}).

\bibitem[{\citenamefont{Sugano et~al.}(1970)\citenamefont{Sugano, Tanabe, and
  Kamimura}}]{SUGANO:70}
\bibinfo{author}{\bibfnamefont{S.}~\bibnamefont{Sugano}},
  \bibinfo{author}{\bibfnamefont{Y.}~\bibnamefont{Tanabe}}, \bibnamefont{and}
  \bibinfo{author}{\bibfnamefont{H.}~\bibnamefont{Kamimura}},
  \emph{\bibinfo{title}{Multiplets of Transition-Metal Ions in Crystal}}
  (\bibinfo{publisher}{Academic}, \bibinfo{address}{New York},
  \bibinfo{year}{1970}).

\bibitem[{\citenamefont{Douglas and Kroll}(1974)}]{DOUGLAS:74}
\bibinfo{author}{\bibfnamefont{M.}~\bibnamefont{Douglas}} \bibnamefont{and}
  \bibinfo{author}{\bibfnamefont{N.~M.} \bibnamefont{Kroll}},
  \bibinfo{journal}{Ann. Phys. (N.Y.)} \textbf{\bibinfo{volume}{82}},
  \bibinfo{pages}{89} (\bibinfo{year}{1974}).

\bibitem[{\citenamefont{Hess}(1986)}]{HESS:86}
\bibinfo{author}{\bibfnamefont{B.~A.} \bibnamefont{Hess}},
  \bibinfo{journal}{Phys.\ Rev.\ A} \textbf{\bibinfo{volume}{33}},
  \bibinfo{pages}{3742} (\bibinfo{year}{1986}).

\bibitem[{\citenamefont{Roos et~al.}(1980)\citenamefont{Roos, Taylor, and
  Siegbahn}}]{ROOS:80}
\bibinfo{author}{\bibfnamefont{B.~O.} \bibnamefont{Roos}},
  \bibinfo{author}{\bibfnamefont{P.~R.} \bibnamefont{Taylor}},
  \bibnamefont{and} \bibinfo{author}{\bibfnamefont{P.~E.~M.}
  \bibnamefont{Siegbahn}}, \bibinfo{journal}{Chem.\ Phys.}
  \textbf{\bibinfo{volume}{48}}, \bibinfo{pages}{157} (\bibinfo{year}{1980}).

\bibitem[{\citenamefont{Siegbahn et~al.}(1980)\citenamefont{Siegbahn, Heiberg,
  Roos, and Levy}}]{SIEGBAHN:80}
\bibinfo{author}{\bibfnamefont{P.~E.~M.} \bibnamefont{Siegbahn}},
  \bibinfo{author}{\bibfnamefont{A.}~\bibnamefont{Heiberg}},
  \bibinfo{author}{\bibfnamefont{B.~O.} \bibnamefont{Roos}}, \bibnamefont{and}
  \bibinfo{author}{\bibfnamefont{B.}~\bibnamefont{Levy}},
  \bibinfo{journal}{Phys.\ Scr.} \textbf{\bibinfo{volume}{21}},
  \bibinfo{pages}{323} (\bibinfo{year}{1980}).

\bibitem[{\citenamefont{Siegbahn et~al.}(1981)\citenamefont{Siegbahn, Heiberg,
  Alml{\"o}f, and Roos}}]{SIEGBAHN:81}
\bibinfo{author}{\bibfnamefont{P.~E.~M.} \bibnamefont{Siegbahn}},
  \bibinfo{author}{\bibfnamefont{A.}~\bibnamefont{Heiberg}},
  \bibinfo{author}{\bibfnamefont{J.}~\bibnamefont{Alml{\"o}f}},
  \bibnamefont{and} \bibinfo{author}{\bibfnamefont{B.~O.} \bibnamefont{Roos}},
  \bibinfo{journal}{J.\ Chem.\ Phys.} \textbf{\bibinfo{volume}{74}},
  \bibinfo{pages}{2384} (\bibinfo{year}{1981}).

\bibitem[{\citenamefont{{K. Andersson, P.-\AA. Malmqvist, B. O. Roos, A. J.
  Sadlej,} and Wolinski}(1990)}]{ANDERSSON:90}
\bibinfo{author}{\bibnamefont{{K. Andersson, P.-\AA. Malmqvist, B. O. Roos, A.
  J. Sadlej,}}} \bibnamefont{and}
  \bibinfo{author}{\bibfnamefont{K.}~\bibnamefont{Wolinski}},
  \bibinfo{journal}{J.\ Phys.\ Chem.} \textbf{\bibinfo{volume}{94}},
  \bibinfo{pages}{5483} (\bibinfo{year}{1990}).

\bibitem[{\citenamefont{{K. Andersson, P.-\AA. Malmqvist} and
  Roos}(1992)}]{ANDERSSON:92}
\bibinfo{author}{\bibnamefont{{K. Andersson, P.-\AA. Malmqvist}}}
  \bibnamefont{and} \bibinfo{author}{\bibfnamefont{B.~O.} \bibnamefont{Roos}},
  \bibinfo{journal}{J.\ Chem.\ Phys.} \textbf{\bibinfo{volume}{96}},
  \bibinfo{pages}{1218} (\bibinfo{year}{1992}).

\bibitem[{\citenamefont{Zaitsevskii and {J.~P.
  Malrieu}}(1995)}]{ZAITSEVSKII:95}
\bibinfo{author}{\bibfnamefont{A.}~\bibnamefont{Zaitsevskii}} \bibnamefont{and}
  \bibinfo{author}{\bibnamefont{{J.~P. Malrieu}}}, \bibinfo{journal}{Chem.\
  Phys.\ Lett.} \textbf{\bibinfo{volume}{233}}, \bibinfo{pages}{597}
  (\bibinfo{year}{1995}).

\bibitem[{\citenamefont{{J. Finley, P.-\AA. Malmqvist, B. O. Roos} and
  Serrano-Andr\'es}(1998)}]{FINLEY:98}
\bibinfo{author}{\bibnamefont{{J. Finley, P.-\AA. Malmqvist, B. O. Roos}}}
  \bibnamefont{and}
  \bibinfo{author}{\bibfnamefont{L.}~\bibnamefont{Serrano-Andr\'es}},
  \bibinfo{journal}{Chem.\ Phys.\ Lett.} \textbf{\bibinfo{volume}{288}},
  \bibinfo{pages}{299} (\bibinfo{year}{1998}).

\bibitem[{\citenamefont{Hess et~al.}(1996)\citenamefont{Hess, Marian, Wahlgren,
  and Gropen}}]{HESS:96}
\bibinfo{author}{\bibfnamefont{B.~A.} \bibnamefont{Hess}},
  \bibinfo{author}{\bibfnamefont{C.~M.} \bibnamefont{Marian}},
  \bibinfo{author}{\bibfnamefont{U.}~\bibnamefont{Wahlgren}}, \bibnamefont{and}
  \bibinfo{author}{\bibfnamefont{O.}~\bibnamefont{Gropen}},
  \bibinfo{journal}{Chem.\ Phys.\ Lett.} \textbf{\bibinfo{volume}{251}},
  \bibinfo{pages}{365} (\bibinfo{year}{1996}).

\bibitem[{\citenamefont{Malmqvist et~al.}(2002)\citenamefont{Malmqvist, Roos,
  and Schimmelpfennig}}]{MALMQVIST:02}
\bibinfo{author}{\bibfnamefont{P.~A.} \bibnamefont{Malmqvist}},
  \bibinfo{author}{\bibfnamefont{B.~O.} \bibnamefont{Roos}}, \bibnamefont{and}
  \bibinfo{author}{\bibfnamefont{B.}~\bibnamefont{Schimmelpfennig}},
  \bibinfo{journal}{Chem.\ Phys.\ Lett.} \textbf{\bibinfo{volume}{357}},
  \bibinfo{pages}{230} (\bibinfo{year}{2002}).

\bibitem[{\citenamefont{Barandiar\'{a}n and Seijo}(1988)}]{BARANDIARAN:88}
\bibinfo{author}{\bibfnamefont{Z.}~\bibnamefont{Barandiar\'{a}n}}
  \bibnamefont{and} \bibinfo{author}{\bibfnamefont{L.}~\bibnamefont{Seijo}},
  \bibinfo{journal}{J.\ Chem.\ Phys.} \textbf{\bibinfo{volume}{89}},
  \bibinfo{pages}{5739} (\bibinfo{year}{1988}).

\bibitem[{\citenamefont{Seijo and Barandiar\'{a}n}(1991)}]{SEIJO:91}
\bibinfo{author}{\bibfnamefont{L.}~\bibnamefont{Seijo}} \bibnamefont{and}
  \bibinfo{author}{\bibfnamefont{Z.}~\bibnamefont{Barandiar\'{a}n}},
  \bibinfo{journal}{J.\ Chem.\ Phys.} \textbf{\bibinfo{volume}{94}},
  \bibinfo{pages}{8158} (\bibinfo{year}{1991}).

\bibitem[{\citenamefont{Barandiar\'{a}n
  et~al.}(2005)\citenamefont{Barandiar\'{a}n, Edelstein, Ordej\'on, Ruip\'erez,
  and Seijo}}]{BARANDIARAN:05:a}
\bibinfo{author}{\bibfnamefont{Z.}~\bibnamefont{Barandiar\'{a}n}},
  \bibinfo{author}{\bibfnamefont{N.~M.} \bibnamefont{Edelstein}},
  \bibinfo{author}{\bibfnamefont{B.}~\bibnamefont{Ordej\'on}},
  \bibinfo{author}{\bibfnamefont{F.}~\bibnamefont{Ruip\'erez}},
  \bibnamefont{and} \bibinfo{author}{\bibfnamefont{L.}~\bibnamefont{Seijo}},
  \bibinfo{journal}{J.\ Solid\ State\ Chem.} \textbf{\bibinfo{volume}{178}},
  \bibinfo{pages}{464} (\bibinfo{year}{2005}).

\bibitem[{\citenamefont{Judd}(1977)}]{JUDD:77}
\bibinfo{author}{\bibfnamefont{B.~R.} \bibnamefont{Judd}},
  \bibinfo{journal}{Phys.\ Rev.\ Lett.} \textbf{\bibinfo{volume}{39}},
  \bibinfo{pages}{242} (\bibinfo{year}{1977}).

\bibitem[{\citenamefont{Morrison}(1980)}]{MORRISON:80:a}
\bibinfo{author}{\bibfnamefont{C.~A.} \bibnamefont{Morrison}},
  \bibinfo{journal}{J.\ Chem.\ Phys.} \textbf{\bibinfo{volume}{72}},
  \bibinfo{pages}{1001} (\bibinfo{year}{1980}).

\bibitem[{\citenamefont{Bettinelli and Moncorg\'e}(2001)}]{BETTINELLI:01}
\bibinfo{author}{\bibfnamefont{M.}~\bibnamefont{Bettinelli}} \bibnamefont{and}
  \bibinfo{author}{\bibfnamefont{R.}~\bibnamefont{Moncorg\'e}},
  \bibinfo{journal}{J.\ Lumin.} \textbf{\bibinfo{volume}{92}},
  \bibinfo{pages}{287} (\bibinfo{year}{2001}).

\bibitem[{\citenamefont{Dorenbos}(2000)}]{DORENBOS:00}
\bibinfo{author}{\bibfnamefont{P.}~\bibnamefont{Dorenbos}},
  \bibinfo{journal}{J.\ Lumin.} \textbf{\bibinfo{volume}{87-89}},
  \bibinfo{pages}{970} (\bibinfo{year}{2000}).

\bibitem[{\citenamefont{Wu et~al.}(2007)\citenamefont{Wu, Gundiah, and
  Cheetman}}]{WU:07:b}
\bibinfo{author}{\bibfnamefont{J.~L.} \bibnamefont{Wu}},
  \bibinfo{author}{\bibfnamefont{G.}~\bibnamefont{Gundiah}}, \bibnamefont{and}
  \bibinfo{author}{\bibfnamefont{A.~K.} \bibnamefont{Cheetman}},
  \bibinfo{journal}{Chem.\ Phys.\ Lett.} \textbf{\bibinfo{volume}{441}},
  \bibinfo{pages}{250} (\bibinfo{year}{2007}).

\bibitem[{\citenamefont{Ganguly et~al.}(1993)\citenamefont{Ganguly, Cheng, and
  O'Neill}}]{GANGULY:93}
\bibinfo{author}{\bibfnamefont{J.}~\bibnamefont{Ganguly}},
  \bibinfo{author}{\bibfnamefont{W.}~\bibnamefont{Cheng}}, \bibnamefont{and}
  \bibinfo{author}{\bibfnamefont{H.~S.~C.} \bibnamefont{O'Neill}},
  \bibinfo{journal}{Am.\ Mineral.} \textbf{\bibinfo{volume}{78}},
  \bibinfo{pages}{583} (\bibinfo{year}{1993}).

\bibitem[{\citenamefont{Armbruster et~al.}(1992)\citenamefont{Armbruster,
  Geiger, and Lager}}]{ARMBRUSTER:92}
\bibinfo{author}{\bibfnamefont{T.}~\bibnamefont{Armbruster}},
  \bibinfo{author}{\bibfnamefont{C.~A.} \bibnamefont{Geiger}},
  \bibnamefont{and} \bibinfo{author}{\bibfnamefont{G.~A.} \bibnamefont{Lager}},
  \bibinfo{journal}{Am.\ Mineral.} \textbf{\bibinfo{volume}{77}},
  \bibinfo{pages}{512} (\bibinfo{year}{1992}).

\bibitem[{\citenamefont{Geiger and Armbruster}(1997)}]{GEIGER:97}
\bibinfo{author}{\bibfnamefont{C.~A.} \bibnamefont{Geiger}} \bibnamefont{and}
  \bibinfo{author}{\bibfnamefont{T.}~\bibnamefont{Armbruster}},
  \bibinfo{journal}{Am.\ Mineral.} \textbf{\bibinfo{volume}{82}},
  \bibinfo{pages}{740} (\bibinfo{year}{1997}).

\bibitem[{\citenamefont{Etschmann et~al.}(2001)\citenamefont{Etschmann,
  Streltsov, Ishizawa, and Maslen}}]{ETSCHMANN:01}
\bibinfo{author}{\bibfnamefont{B.}~\bibnamefont{Etschmann}},
  \bibinfo{author}{\bibfnamefont{V.}~\bibnamefont{Streltsov}},
  \bibinfo{author}{\bibfnamefont{N.}~\bibnamefont{Ishizawa}}, \bibnamefont{and}
  \bibinfo{author}{\bibfnamefont{E.~N.} \bibnamefont{Maslen}},
  \bibinfo{journal}{Acta\ Crystallogr.\ B} \textbf{\bibinfo{volume}{57}},
  \bibinfo{pages}{136} (\bibinfo{year}{2001}).

\bibitem[{\citenamefont{Nakatsuka et~al.}(1995)\citenamefont{Nakatsuka,
  Yoshiasa, and Takeno}}]{NAKATSUKA:95}
\bibinfo{author}{\bibfnamefont{A.}~\bibnamefont{Nakatsuka}},
  \bibinfo{author}{\bibfnamefont{A.}~\bibnamefont{Yoshiasa}}, \bibnamefont{and}
  \bibinfo{author}{\bibfnamefont{S.}~\bibnamefont{Takeno}},
  \bibinfo{journal}{Acta\ Crystallogr.\ B} \textbf{\bibinfo{volume}{51}},
  \bibinfo{pages}{737} (\bibinfo{year}{1995}).

\bibitem[{\citenamefont{Patzke et~al.}(1999)\citenamefont{Patzke, Wartchow, and
  Binnewies}}]{PATZKE:99}
\bibinfo{author}{\bibfnamefont{G.}~\bibnamefont{Patzke}},
  \bibinfo{author}{\bibfnamefont{R.}~\bibnamefont{Wartchow}}, \bibnamefont{and}
  \bibinfo{author}{\bibfnamefont{M.}~\bibnamefont{Binnewies}},
  \bibinfo{journal}{Z.\ Kristallogr.} \textbf{\bibinfo{volume}{214}},
  \bibinfo{pages}{143} (\bibinfo{year}{1999}).

\bibitem[{\citenamefont{Sawada}(1997)}]{SAWADA:97}
\bibinfo{author}{\bibfnamefont{H.}~\bibnamefont{Sawada}}, \bibinfo{journal}{J.\
  Solid\ State\ Chem.} \textbf{\bibinfo{volume}{132}}, \bibinfo{pages}{300}
  (\bibinfo{year}{1997}).

\end{thebibliography}

\clearpage
\begin{table}[h]
\caption{
        Definitions of the $D_2$ totally symmetric displacements of the CeO$_8$ moiety chosen in this work:
        Stretching, bending, and twisting $O_h$ symmetry coordinates.
        The labels of the oxygen atoms and the chosen cartesian axes 
        are defined in Fig.~\protect\ref{FIG:S1-S6}:
        In the reference cube,
        the symmetry independent oxygen atoms $a_1$ and $e_1$ are located 
        at $(\sqrt{2},0,1)d/\sqrt{3}$ and $(0,\sqrt{2},1)d/\sqrt{3}$ respectively,
        with $d$ being the Ce-O distance.
        $\hat{R}$ is the $D_2$ symmetrization operator: 
        the normalized addition of the group symmetry operations 
        (the identity and the three $180^{\circ}$ rotations around the cartesian axes).
        $\delta x_{a1}$ is the $x$ cartesian displacement of the Oxygen atom $a1$
        from its position in the reference cube:
        $\delta x_{a1}=x_{a1}-x_{a1,ref}$;
        identical definitions stand for the other cartesian displacements.
                }
\label{TAB:S1-S6}
\begin{ruledtabular}
\begin{tabular}{ccc}
 \multicolumn{3}{c}{
$\hat{R} =  \frac{1}{2} \left(  \hat{I} + \hat{C}_{2x} + \hat{C}_{2y} + \hat{C}_{2z} \right)$
}
\\ Displacement & $O_h$ irrep & Definition
\\ symmetric stretching & $a_{1g}$ &
$S_1 = 
\frac{1}{\sqrt{6}} \,\hat{R}\,
\left[ \left( \sqrt{2}\,\delta x_{a1} + \delta z_{a1} \right) + \left( \sqrt{2}\,\delta y_{e1} + \delta z_{e1} \right) \right] $
\\ asymmetric stretching &  $e_{g}\epsilon$ &
$S_2 = 
\frac{1}{\sqrt{6}} \,\hat{R}\,
\left[ \left( \sqrt{2}\,\delta x_{a1} + \delta z_{a1} \right) - \left( \sqrt{2}\,\delta y_{e1} + \delta z_{e1} \right) \right] $
\\ symmetric bending & $e_{g}\theta$ &
$S_3 = 
\frac{1}{\sqrt{6}} \,\hat{R}\,
\left[ \left( -\delta x_{a1} + \sqrt{2}\,\delta z_{a1} \right) + \left( -\delta y_{e1} + \sqrt{2}\,\delta z_{e1} \right) \right] $
\\ asymmetric bending & $e_{g}\epsilon$ &
$S_4 = 
\frac{1}{\sqrt{6}} \,\hat{R}\,
\left[ \left( -\delta x_{a1} + \sqrt{2}\,\delta z_{a1} \right) - \left( -\delta y_{e1} + \sqrt{2}\,\delta z_{e1} \right) \right] $
\\ symmetric twisting & $e_{u}\theta$ &
$S_5 = 
\frac{1}{\sqrt{2}} \,\hat{R}\,
\left[  -\delta y_{a1} + \delta x_{e1}  \right] $
\\ asymmetric twisting & $e_{u}\epsilon$ &
$S_6 = 
\frac{1}{\sqrt{2}} \,\hat{R}\,
\left[  \delta y_{a1} + \delta x_{e1}  \right] $
\\
 \end{tabular}
\end{ruledtabular}
\end{table}
\clearpage
\begin{sidewaystable}[h]
\caption{Crystallographic data of some garnets:
              lattice constant $a$ and special position (d)
              of the $Ia\overline{3}d$ (230) space group $x, y, z$.
              Local $D_2$ 8-fold oxygen coordination of the fixed position (c) occupied by Ce as a dopant:
              cation-oxygen distance in the reference cube $d_{ref}$
              and $D_2$ displacements $S_1$ - $S_6$ as defined in Table~\ref{TAB:S1-S6};
              $S_1$ is defined with respect to a cube with cation-oxygen distance $d=$2.34~\AA\ 
              (note that the use of $d_{ref}$, $S_1=0$, and $S_2 - S_6$ or
              $d_{ref}=$2.34~\AA, and  $S_1 - S_6$ are two valid alternatives to define the oxygen coordinations).
              $a$, $d_{ref}$ and $S_1$ - $S_6$ are given in \AA\ and
              $x, y, z$  in fractional cell units.
              The lowest $4f \rightarrow 5d$ transitions
              calculated for the \CeOviiimxiii\  clusters,
              embedded in a common cubic confinement potential and having 
              the local structures of the undistorted garnets, are given in the last column in \cmm1;
              its stabilization with respect to a reference cube of $d=$2.34~\AA\  is given in parenthesis.
             }
\label{TAB:garnets}
\renewcommand{\arraystretch}{0.64}   
\begin{ruledtabular}
\begin{tabular}{ccccccccccccccc}
Garnet 		& 							& Ref. & $a$ & $x$ & $y$ & $z$ & $d_{ref}$ & $S_1$ & $S_2$ & $S_3$ & $S_4$ & $S_5$ & $S_6$ & $4f \rightarrow 5d$  \\ \hline
Pyrope		& Mg$_3$Al$_2$Si$_3$O$_{12}$ 	& \onlinecite{GANGULY:93}		& 11.4566 & -0.03290 & 0.05010 & 0.15280 & 2.243 & -0.2731 & -0.2191 & -0.0768 & -0.1215 & 0.9900 & -0.0892 & 22750~(-3250) \\
Almandine 	& Fe$_3$Al$_2$Si$_3$O$_{12}$ 	& \onlinecite{ARMBRUSTER:92} 	& 11.5250 & -0.03401 & 0.04901 & 0.15278 & 2.267 & -0.2070 & -0.2334 & -0.1126 & -0.1279 & 1.0084 & -0.1028 & 23380~(-2620) \\
Spessartine	& Mn$_3$Al$_2$Si$_3$O$_{12}$ 	& \onlinecite{GEIGER:97} 		& 11.6190 & -0.03491 & 0.04791 & 0.15250 & 2.295 & -0.1278 & -0.2490 & -0.1414 & -0.1337 & 1.0281 & -0.1243 & 24100~(-1900) \\
Grossular 		& Ca$_3$Al$_2$Si$_3$O$_{12}$ 	& \onlinecite{GANGULY:93}		& 11.8515 & -0.03760 & 0.04540 & 0.15130 & 2.368 &  0.0790 & -0.2752 & -0.2203 & -0.1378 & 1.0725 & -0.1909 & 25670~(\,\,\,-330) \\
LuAG 		& Lu$_3$Al$_5$O$_{12}$ 	    		& \onlinecite{EULER:65}		& 11.9060 & -0.02940 & 0.05370 & 0.15090 & 2.303 & -0.1042 & -0.1683 &  0.0555 & -0.0866 & 0.9766 & -0.1044 & 24910~(-1090) \\
YbAG 		& Yb$_3$Al$_5$O$_{12}$ 	    		& \onlinecite{EULER:65}		& 11.9310 & -0.02960 & 0.05290 & 0.15040 & 2.313 & -0.0759 & -0.1819 &  0.0497 & -0.0906 & 0.9858 & -0.1286 & 25140~(\,\,\,-860) \\		
			& Lu$_2$CaMg$_2$Si$_3$O$_{12}$ 	& \onlinecite{SETLUR:06}		& 11.9750 & -0.03510 & 0.05380 & 0.15780 & 2.333 & -0.0200 & -0.1522 & -0.1619 & -0.1139 & 1.0153 &  0.0956 & 25240~(\,\,\,-760) \\
ErAG 		& Er$_3$Al$_5$O$_{12}$ 			& \onlinecite{ETSCHMANN:01}	& 11.9928 & -0.03039 & 0.05124 & 0.14915 & 2.339 & -0.0039 & -0.2047 &  0.0297 & -0.0944 & 1.0046 & -0.1854 & 25710~(\,\,\,-290) \\
YAG 			& Y$_3$Al$_5$O$_{12}$ 			& \onlinecite{EULER:65}		& 12.0000 & -0.03060 & 0.05120 & 0.15000 & 2.339 & -0.0019 & -0.2094 &  0.0156 & -0.1018 & 1.0099 & -0.1613 & 25680~(\,\,\,-320) \\
Andradite 		& Ca$_3$Fe$_2$Si$_3$O$_{12}$ 	& \onlinecite{QUARTIERI:06}	& 12.0578 & -0.03940 & 0.04870 & 0.15540 & 2.398 &  0.1648 & -0.2102 & -0.2895 & -0.1260 & 1.0718 & -0.0351 & 26210~(\,\,+210) \\
GdAG 		& Gd$_3$Al$_5$O$_{12}$ 			& \onlinecite{EULER:65}		& 12.1130 & -0.03110 & 0.05090 & 0.14900 & 2.368 &  0.0788 & -0.2060 &  0.0101 & -0.0932 & 1.0180 & -0.1957 & 26340~(\,\,+340) \\
LuGG  		& Lu$_3$Ga$_5$O$_{12}$ 			& \onlinecite{EULER:65}		& 12.1880 & -0.02520 & 0.05700 & 0.15060 & 2.322 & -0.0520 & -0.1378 &  0.1961 & -0.0727 & 0.9580 & -0.0755 & 25870~(\,\,\,-130) \\
YbGG  		& Yb$_3$Ga$_5$O$_{12}$ 		& \onlinecite{EULER:65}		& 12.2040 & -0.02590 & 0.05630 & 0.15190 & 2.328 & -0.0336 & -0.1561 &  0.1596 & -0.0899 & 0.9744 & -0.0459 & 25860~(\,\,\,-140) \\
  			& Ca$_3$Sc$_2$Si$_3$O$_{12}$ 	& \onlinecite{QUARTIERI:06}	& 12.2500 & -0.04004 & 0.05010 & 0.15887 & 2.427 &  0.2454 & -0.1972 & -0.3362 & -0.1401 & 1.0894 &  0.0842 & 26580~(\,\,+580) \\
YGG  		& Y$_3$Ga$_5$O$_{12}$ 			& \onlinecite{NAKATSUKA:95}	& 12.2730 & -0.02740 & 0.05460 & 0.14930 & 2.363 &  0.0647 & -0.1629 &  0.1338 & -0.0757 & 0.9876 & -0.1440 & 26640~(\,\,+640) \\
HoGG  		& Ho$_3$Ga$_5$O$_{12}$ 		& \onlinecite{PATZKE:99}		& 12.2900 & -0.02740 & 0.05520 & 0.15020 & 2.362 &  0.0616 & -0.1550 &  0.1293 & -0.0771 & 0.9861 & -0.1101 & 26610~(\,\,+610) \\
DyGG 		& Dy$_3$Ga$_5$O$_{12}$ 		& \onlinecite{PATZKE:99}		& 12.3060 & -0.02780 & 0.05490 & 0.14990 & 2.369 &  0.0823 & -0.1558 &  0.1194 & -0.0752 & 0.9896 & -0.1229 & 26740~(\,\,+740) \\
TbGG 		& Tb$_3$Ga$_5$O$_{12}$ 		& \onlinecite{SAWADA:97}		& 12.3474 & -0.02820 & 0.05470 & 0.14950 & 2.381 &  0.1166 & -0.1538 &  0.1115 & -0.0712 & 0.9933 & -0.1377 & 26980~(\,\,+980) \\
GdGG 		& Gd$_3$Ga$_5$O$_{12}$ 		& \onlinecite{SAWADA:97}		& 12.3829 & -0.02890 & 0.05420 & 0.14940 & 2.394 &  0.1538 & -0.1578 &  0.0901 & -0.0719 & 1.0019 & -0.1473 & 27180~(+1180) \\
SmGG 		& Sm$_3$Ga$_5$O$_{12}$ 		& \onlinecite{SAWADA:97}		& 12.4361 & -0.02920 & 0.05300 & 0.14900 & 2.412 &  0.2031 & -0.1816 &  0.0777 & -0.0818 & 1.0190 & -0.1748 & 27410~(+1410) \\
\end{tabular}
\end{ruledtabular}
\end{sidewaystable}
%
%
\clearpage
\section*{Figure captions}
\begin{figure}[ht]
\caption{
              $D_2$ totally symmetric displacements of the CeO$_8$ moiety:
              Stretching, bending, and twisting $O_h$ symmetry coordinates 
              $S_1$ ($a_{1g}$ symmetric stretching -breathing-),
              $S_2$ ($e_{g}\epsilon$ asymmetric stretching), 
              $S_3$ ($e_{g}\theta$ symmetric bending), 
              $S_4$ ($e_{g}\epsilon$ asymmetric bending), 
              $S_5$ ($e_{u}\theta$ symmetric twisting), and
              $S_6$ ($e_{u}\epsilon$ asymmetric twisting).
              See Table~\ref{TAB:S1-S6} for the detailed definitions.
             }
\label{FIG:S1-S6}
\end{figure}
\begin{figure}[ht]
\caption{
              $4f$ (dashed lines) and $5d$ (full lines) energy levels 
              of the  \protect\CeOviiimxiii\  embedded cluster (relative to the ground state)
              as a function of the $S_1$ - $S_6$ $D_2$ oxygen displacement coordinates,
              calculated without spin-orbit coupling.
              The differences between the $5d$ and $4f$ energy centroids (dot-dashed lines)
              and between the ligand field stabilization energies of the  $5d$ and $4f$  lowest
              levels  (doted lines) are also shown;
              the lowest $4f \rightarrow 5d$ transition equals the subtraction of these two.
              Experimental values of $S_1$ - $S_6$ of 21 pure garnets are shown as small vertical lines.  
             }
\label{FIG:S135S246-spinfree}
\end{figure}
\begin{figure}[ht]
\caption{
              $4f$  and $5d$  energy levels 
              of the  \protect\CeOviiimxiii\  embedded cluster 
              as a function of  $S_1$ - $S_6$,
              calculated with spin-orbit coupling.
              All levels are $D_2^\prime$ $\Gamma_5$ Kramer doublets.
              See Fig.\protect\ref{FIG:S135S246-spinfree} caption.
             }
\label{FIG:S135S246-spinorbit}
\end{figure}
\begin{figure}[ht]
\caption{
             $4f$  energy levels of the  \protect\CeOviiimxiii\  embedded cluster 
             as a function of  $S_1$ - $S_6$, calculated without spin-orbit coupling.
             Dotted lines: $^2A$ levels; 
             full lines: $^2B_1$ levels; 
             dashed lines: $^2B_2$ and $^2B_3$ levels.              
             Experimental values of $S_1$ - $S_6$ of 21 pure garnets are shown as small vertical lines.  
             }
\label{FIG:S135S246-spinfree-4f}
\end{figure}
\begin{figure}[ht]
\caption{
             $4f$  energy levels of the  \protect\CeOviiimxiii\  embedded cluster 
             as a function of  $S_1$ - $S_6$, calculated with spin-orbit coupling.
             All levels are $D_2^\prime$ $\Gamma_5$ Kramer doublets.              
             Experimental values of $S_1$ - $S_6$ of 21 pure garnets are shown as small vertical
             lines.               
             }
\label{FIG:S135S246-spinorbit-4f}
\end{figure}
\clearpage
 \begin{center}
 \vfill
 \resizebox{15cm}{!}{ \rotatebox{0}{\includegraphics{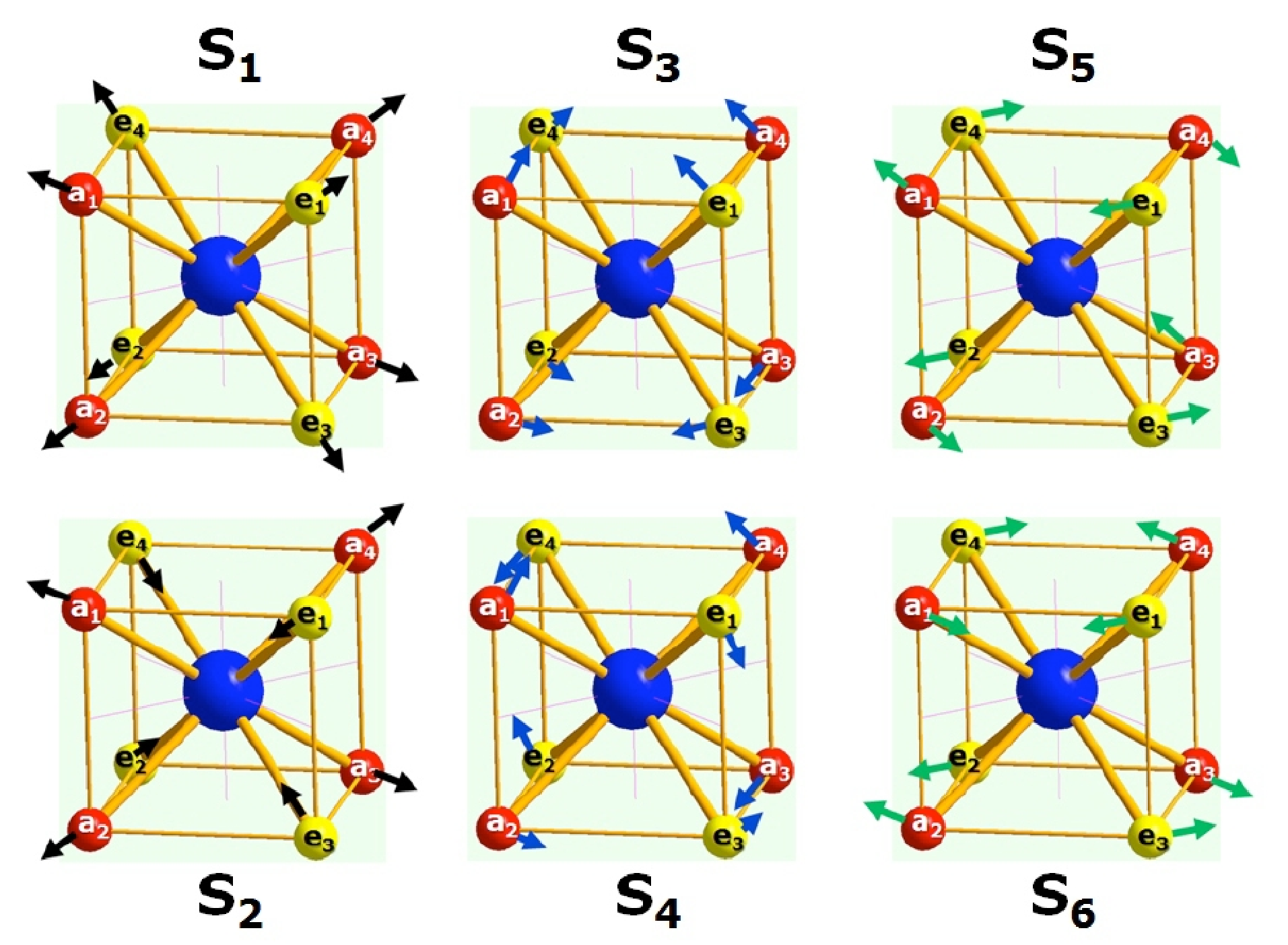}}}
 \vfill
Figure~\ref{FIG:S1-S6}. Seijo and Barandiar\'an
 \end{center}
\thispagestyle{empty}
\clearpage
 \begin{center}
 \vfill
 \resizebox{16cm}{!}{ \rotatebox{-90}{\includegraphics{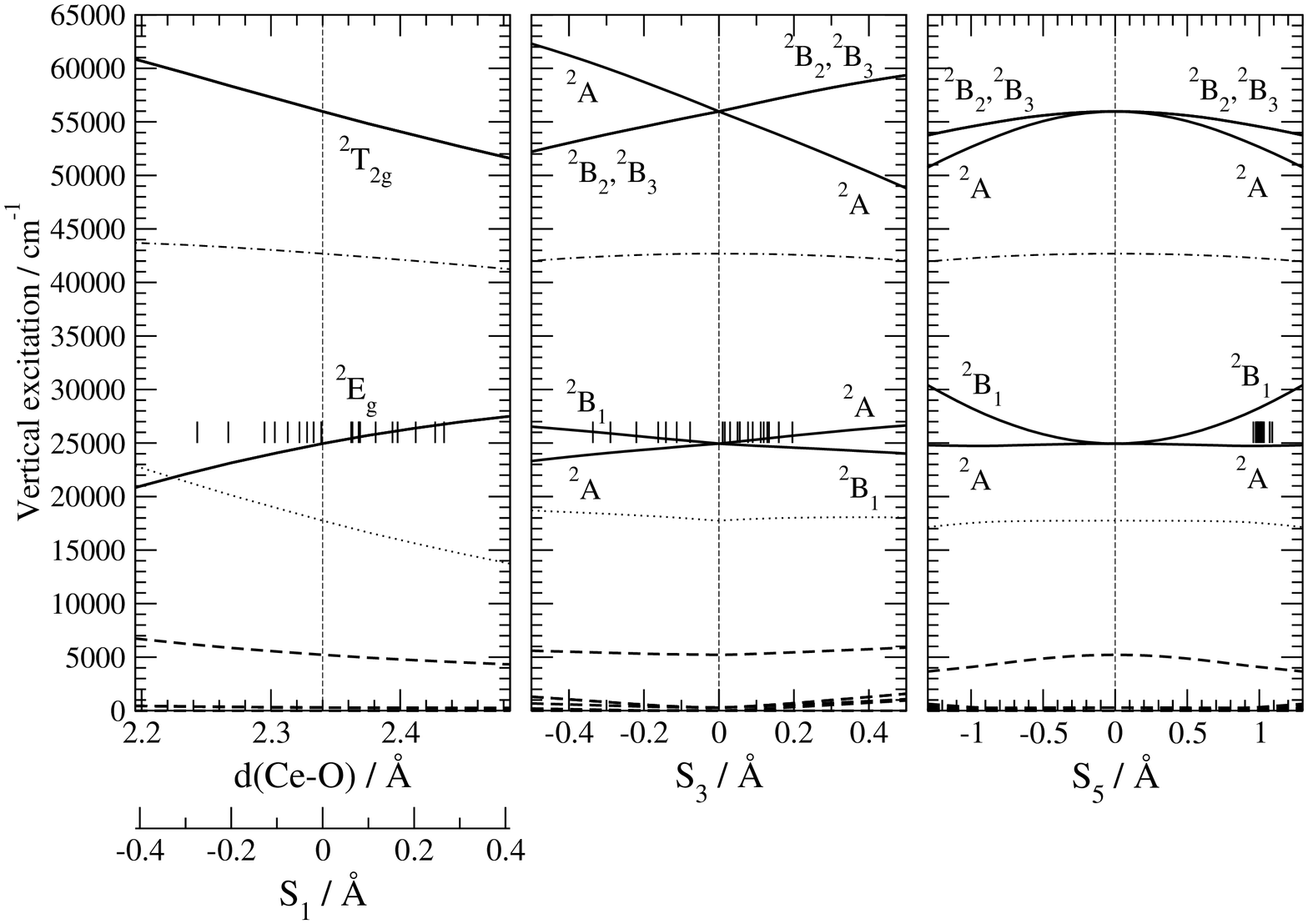}}}\\[-6mm]
 \resizebox{16cm}{!}{ \rotatebox{-90}{\includegraphics{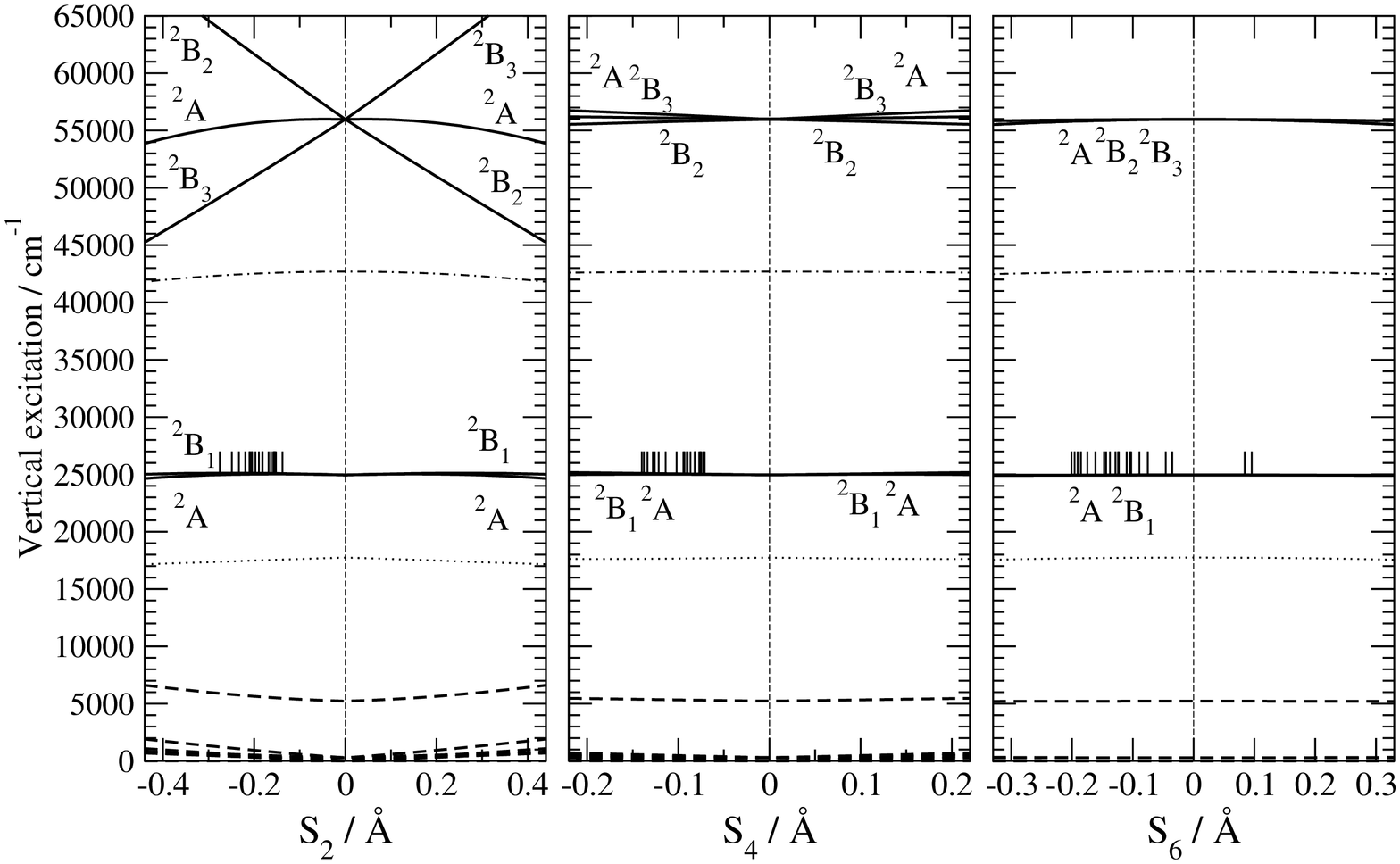}}}
 \vfill
Figure~\ref{FIG:S135S246-spinfree}. Seijo and Barandiar\'an
 \end{center}
\thispagestyle{empty}
\clearpage
 \begin{center}
 \vfill
 \resizebox{16cm}{!}{ \rotatebox{-90}{\includegraphics{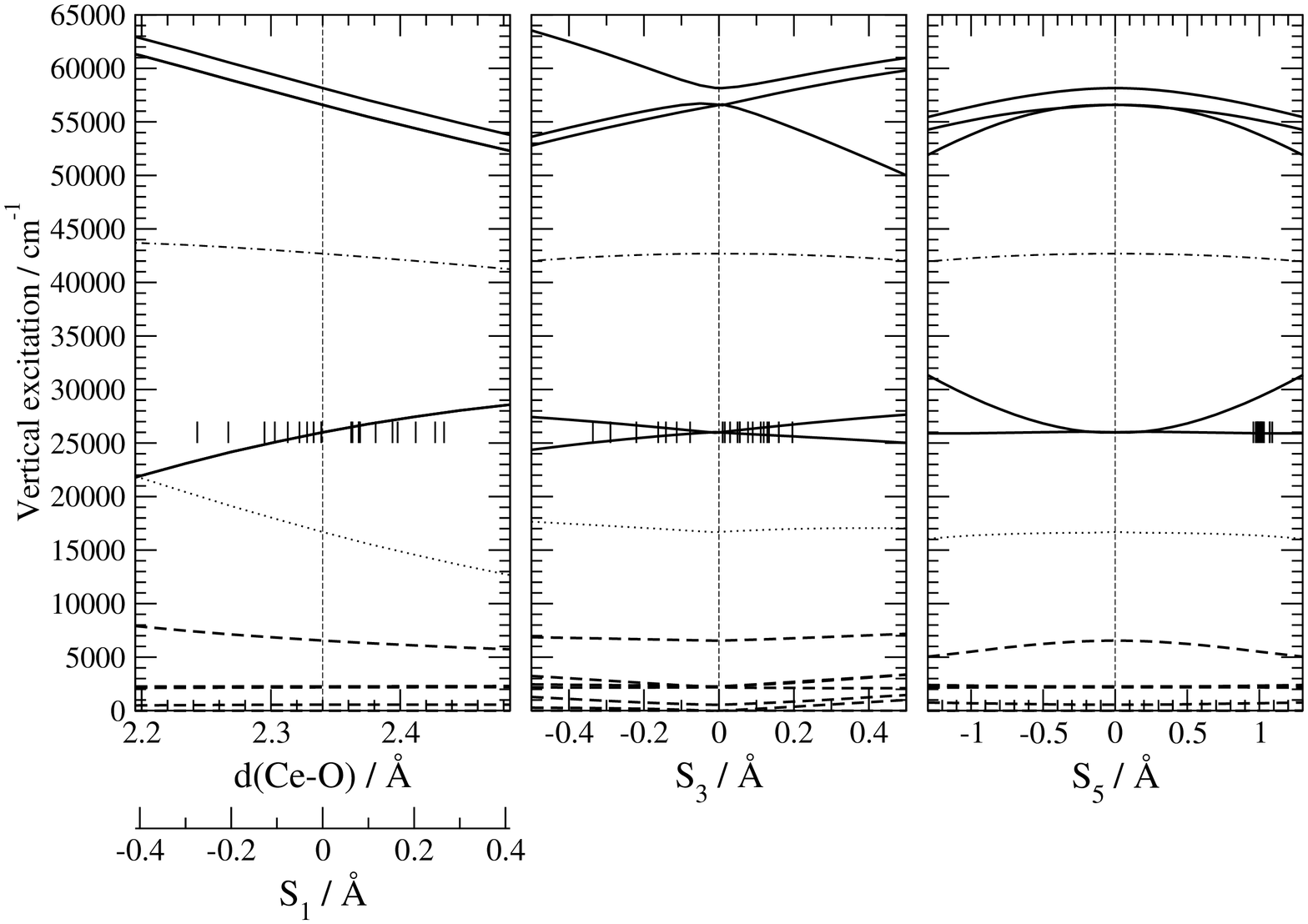}}}\\[-6mm]
 \resizebox{16cm}{!}{ \rotatebox{-90}{\includegraphics{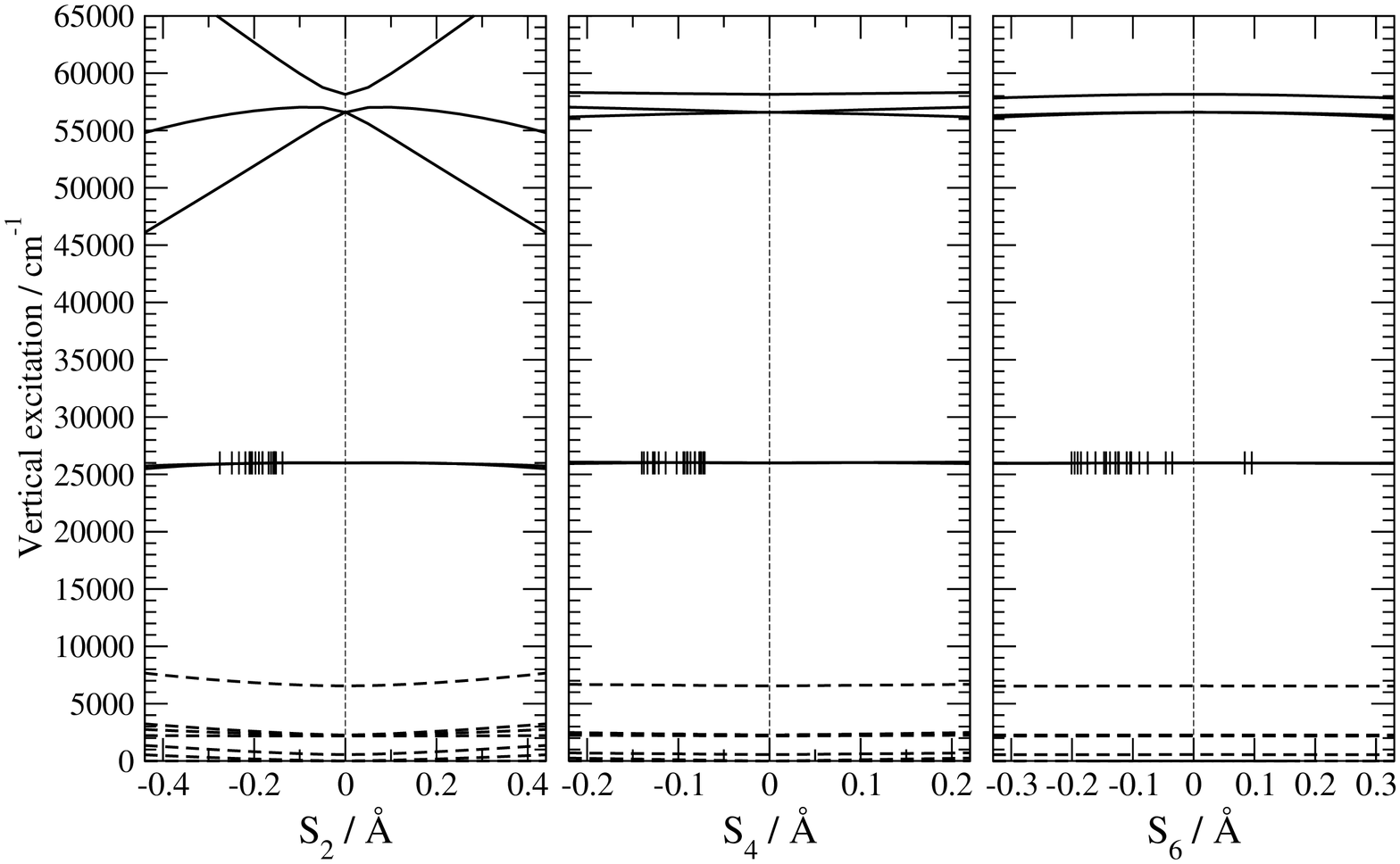}}}
 \vfill
Figure~\ref{FIG:S135S246-spinorbit}. Seijo and Barandiar\'an
 \end{center}
\thispagestyle{empty}
\clearpage
 \begin{center}
 \vfill
 \resizebox{16cm}{!}{ \rotatebox{-90}{\includegraphics{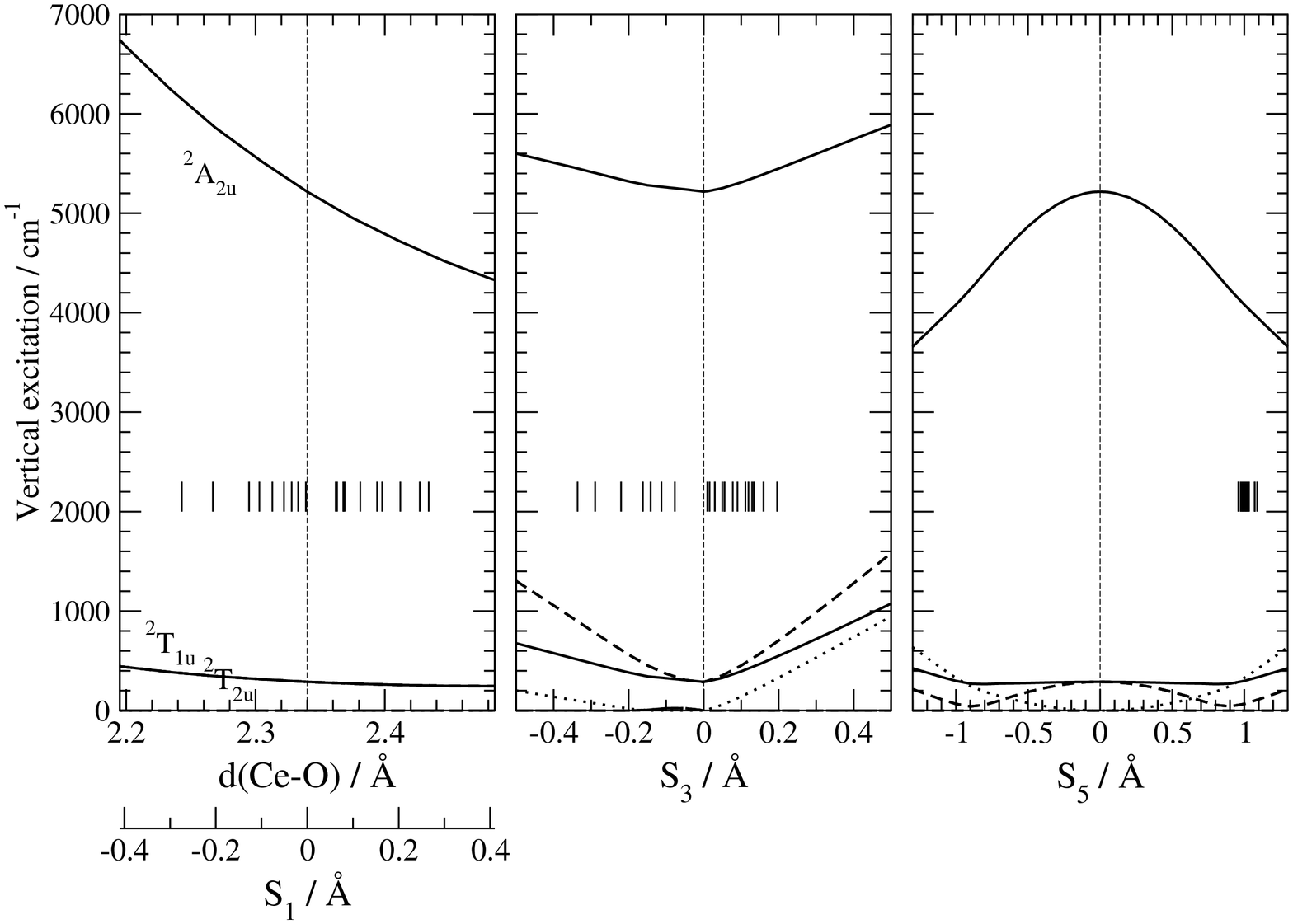}}}\\[-6mm]
 \resizebox{16cm}{!}{ \rotatebox{-90}{\includegraphics{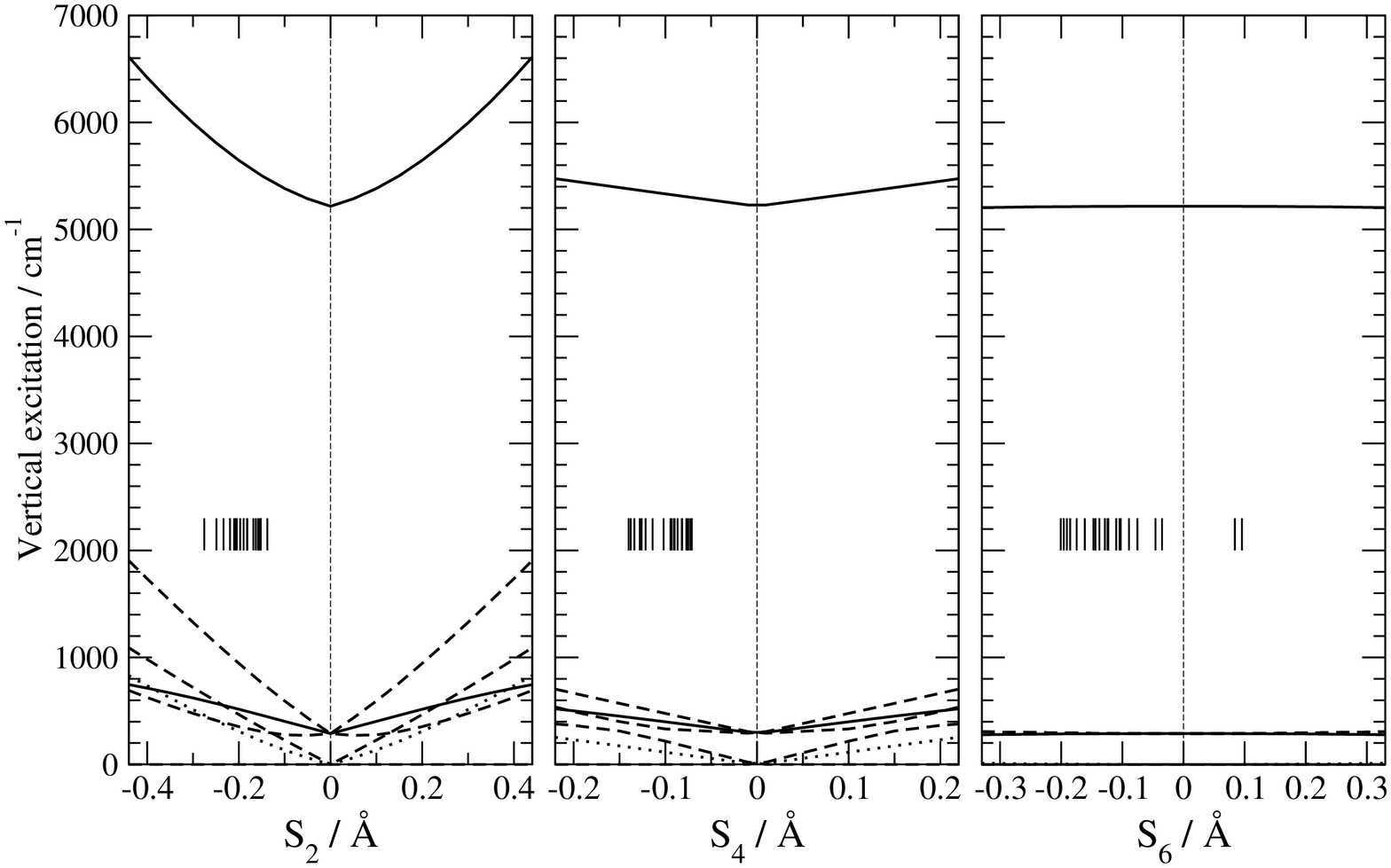}}}
 \vfill
Figure~\ref{FIG:S135S246-spinfree-4f}. Seijo and Barandiar\'an
 \end{center}
\thispagestyle{empty}
\clearpage
 \begin{center}
 \vfill
 \resizebox{16cm}{!}{ \rotatebox{-90}{\includegraphics{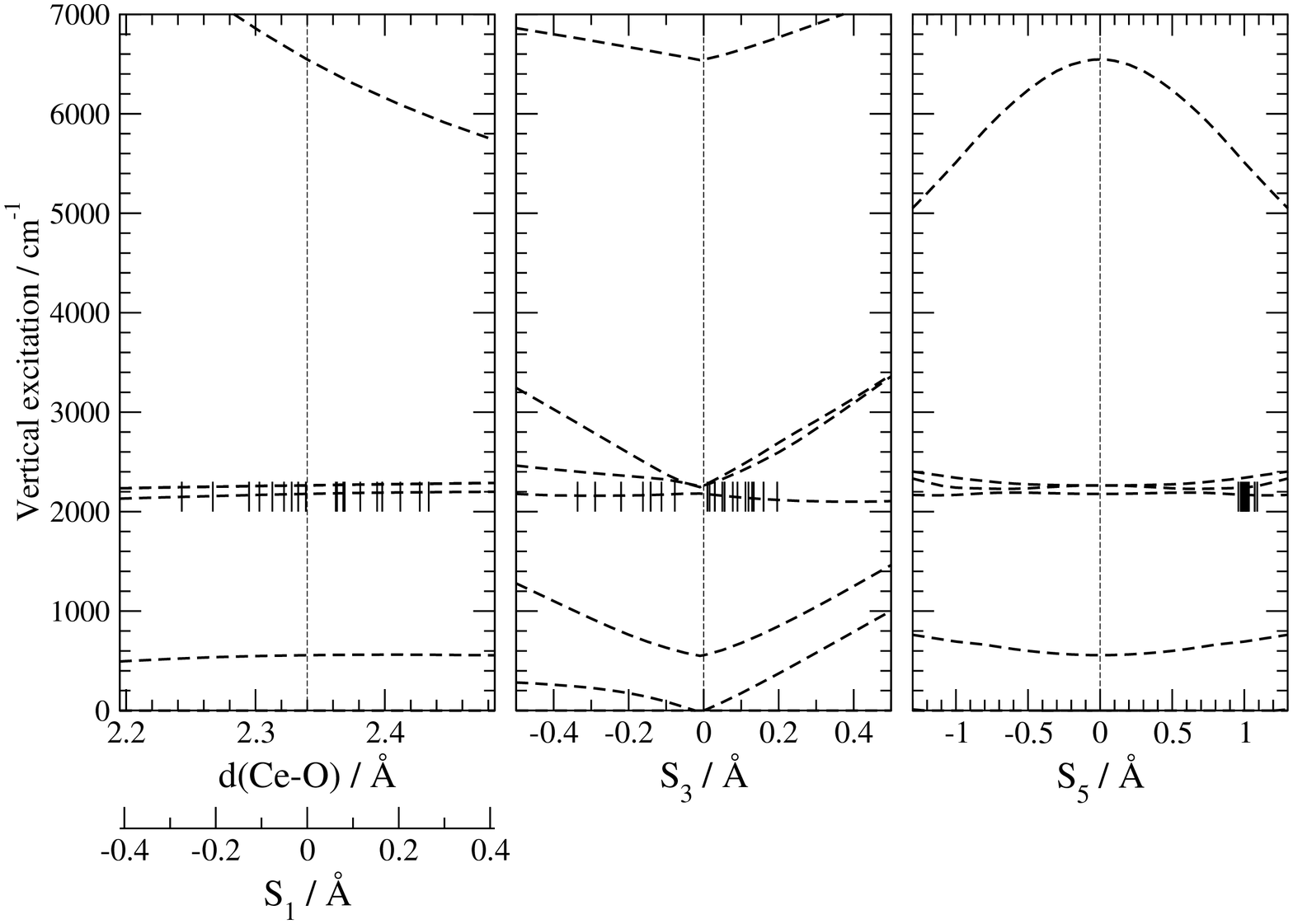}}}\\[-6mm]
 \resizebox{16cm}{!}{ \rotatebox{-90}{\includegraphics{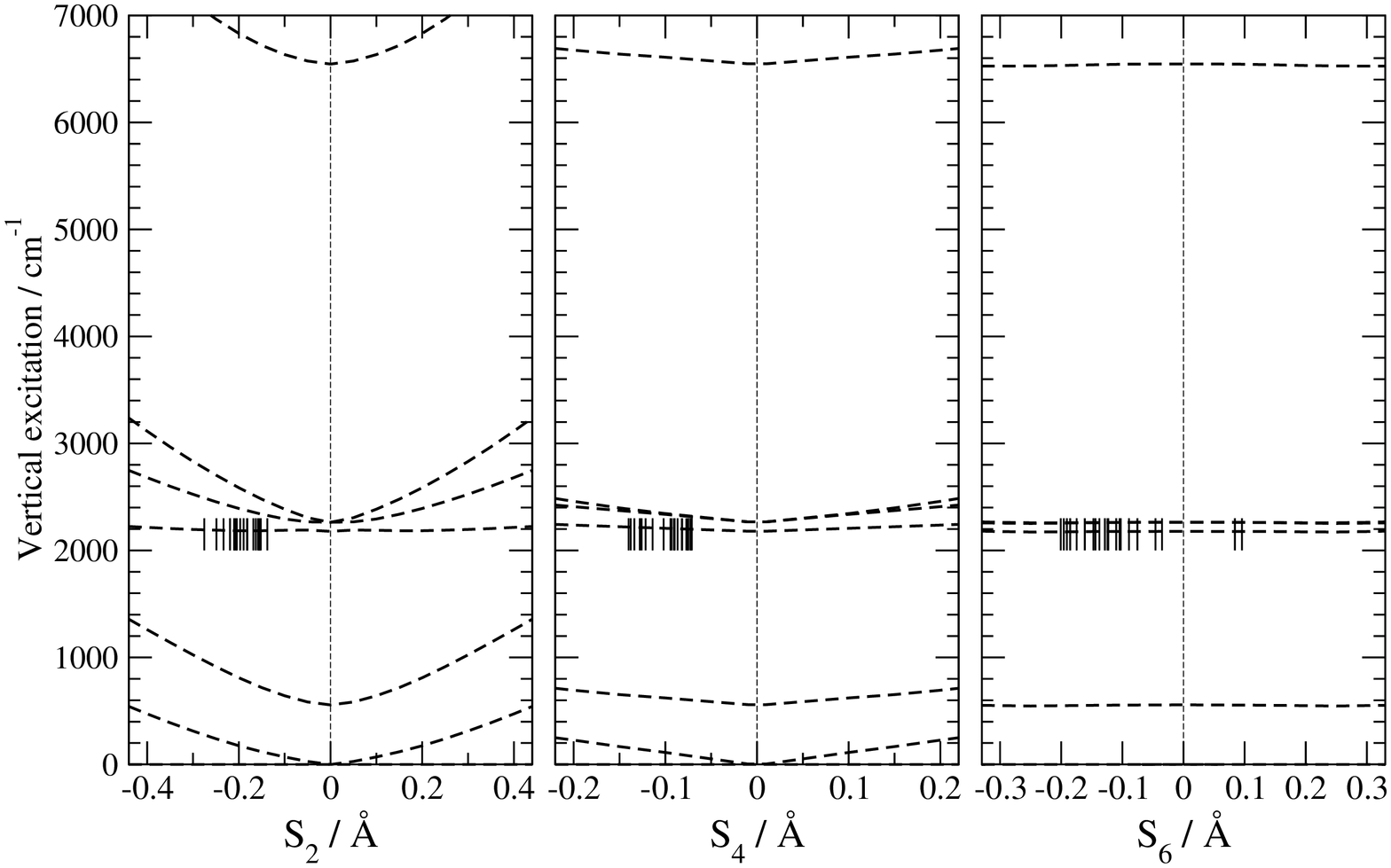}}}
 \vfill
Figure~\ref{FIG:S135S246-spinorbit-4f}. Seijo and Barandiar\'an
 \end{center}
\thispagestyle{empty}

\end{document}